\documentclass[useAMS]{mn2e}
\usepackage{graphicx}
\usepackage[authoryear]{natbib}

\title[Read-out streaks in the UVOT]{The use and calibration of read-out streaks to 
increase the dynamic range of the {\em Swift} Ultraviolet/Optical Telescope}

\author[M.J. Page et al.]
{M.J. Page$^{1}$, N.P.M. Kuin$^{1}$, 
A.A. Breeveld$^{1}$, 
B. Hancock$^{1}$, 
S.T. Holland$^{2}$,
\and 
F.E. Marshall$^{3}$,
S. Oates$^{1}$, 
P.W.A. Roming$^{4,5}$,
M.H. Siegel$^{5}$,
P.J. Smith$^{1}$, 
\and
M. Carter$^{1}$, 
M. De Pasquale$^{1}$,
M. Symeonidis$^{1}$, 
V. Yershov$^{1}$, 
A.P. Beardmore$^{6}$\\ 
\\
$^{1}$Mullard Space Science Laboratory, University College London,
Holmbury St Mary, Dorking, Surrey, RH5 6NT, UK\\
$^{2}$Space Telescope Science Centre, 3700 San Martin Drive, Baltimore, 
MD 21218, USA\\
$^{3}$Astrophysics Science Division, Code 660.1, Goddard Space Flight 
Center, 8800 Greenbelt Road, Greenbelt, MA 20771, USA\\
$^{4}$Space Science \& Engineering Division, Southwest Research Institute, 
P.O. Drawer 28510, San Antonio, TX 78228-0510, USA\\
$^{5}$Department of Astronomy \& Astrophysics, Penn State University, 
525 Davey Laboratory, University Park, PA 16802, USA\\
$^{6}$Dept. of Physics and Astronomy, University of Leicester, 
Leicester, LE1 7RH, UK\\
}

\begin{document}

\date{Accepted ----. Received ----; in original form ----}

\pagerange{\pageref{firstpage}--\pageref{lastpage}} 
\pubyear{2013}
\maketitle

\label{firstpage}

\begin{abstract}
  The dynamic range of photon counting micro-channel-plate (MCP)
  intensified charged-coupled device (CCD) instruments such as the
  {\em Swift} Ultraviolet/Optical Telescope (UVOT) and the XMM-Newton
  Optical Monitor (XMM-OM) is limited at the bright end by coincidence
  loss, the superposition of multiple photons in the individual frames
  recorded by the CCD. Photons which arrive during the brief period in
  which the image frame is transferred for read out of the CCD are 
  displaced in the transfer direction in the recorded images. 
  For sufficiently bright sources,
  these displaced counts form read-out streaks. Using UVOT observations of
  Tycho-2 stars, we investigate the use
  of these read-out streaks to obtain photometry for sources which are
  too bright (and hence have too much coincidence loss) for normal
  aperture photometry to be reliable. For
  read-out-streak photometry, the bright-source limiting factor is
  coincidence loss within the MCPs rather than the CCD. We find that 
photometric measurements can be
  obtained for stars up to 2.4 magnitudes brighter than the usual
  full-frame coincidence-loss limit by using the read-out streaks. 
  The resulting bright-limit Vega magnitudes in the UVOT passbands are 
  UVW2=8.80, UVM2=8.27, UVW1=8.86, u=9.76, b=10.53, v=9.31 and White=11.71; 
  these limits are independent of the windowing mode of the camera.
  We find that a
  photometric precision of 0.1 mag can be achieved through read-out
  streak measurements. A suitable method for the measurement of
  read-out streaks is described and all necessary calibration factors
  are given.
\end{abstract}


\section{Introduction}
\label{sec:introduction}

The Ultraviolet/Optical Telescope \citep[UVOT;][]{roming05} is a 30~cm
optical/UV telescope mounted on the NASA {\em Swift} observatory
\citep{gehrels04}.  The UVOT is coaligned with the X-ray Telescope
\citep[XRT;][]{burrows05} on a rapidly-repointing satellite bus so that
the two instruments can observe the afterglows of gamma-ray bursts
within two minutes of a gamma-ray burst being detected by the
wide-field Burst Alert Telescope \citep[BAT;][]{barthelmy05}.

The UVOT is of a modified Ritchey Chr\'{e}tien design. After
reflections on the primary and secondary mirrors, the incoming light
passes through a hole in the primary to a flat tertiary mirror which
directs the light to one of two identical filter wheel and detector chains.
There are seven imaging filters mounted in the UVOT filter wheel
together with two grisms for low-dispersion spectroscopy. Three optical filters,
u, b and v, cover similar wavelength ranges to the Johnson UBV set
\citep{johnson51}, and three UV filters, UVW2, UVM2 and UVW1, have central
wavelengths of 1928\AA, 2246\AA\ and 2600\AA\ respectively. The remaining 
filter, WHITE, transmits over the full UVOT bandpass
(1600--8000\AA) to maximise throughput. 

The UVOT detector is a micro-channel plate (MCP) intensified charge
coupled device (CCD) \citep[MIC;][]{fordham89}.  Individual incident photons
liberate electrons in a multi-alkali (S20) photo-cathode, which are
multiplied a million-fold using MCPs in series. These electrons strike
a phosphor screen, producing photons which are fed to the CCD via a
fibre-optic taper. The resulting cascade of photons arriving on the
CCD is centroided on-board to a precision of one eighth of a CCD
pixel. The $256 \times 256$-pixel science area of the 
CCD thus localises individual
incoming photons on a grid of $2048 \times 2048$ pixels of size 
$0.5$~arcsec~$\times$~$0.5$~arcsec with a time resolution equal to the frame
time of the CCD (usually 11.0329 ms). The onboard centroiding produces a
low-level modulo-8 fixed-pattern distortion \citep{kawakami94} which
is routinely corrected in the ground processing.  Data are recorded
and transmitted to the ground as an event list, in which the arrival
times and positions of individual photons are recorded, or as an image
accumulated over a timed exposure. To reduce data volume, images are
often binned by a factor of 2 in $x$ and $y$ before transmission to
the ground.

The {\em Swift} UVOT has an almost-identical optical design and a similar 
detector system to the
{\em XMM-Newton} Optical Monitor \citep[XMM-OM][]{mason01}.  The XMM-OM differs
from the UVOT mainly in its UV throughput (which is lower in XMM-OM), its
control and data processing computers (which are of an earlier vintage in
XMM-OM) and in its operating modes.

The detector response is asymptotically linear
when the arrival rate of photons is small 
compared to the
CCD readout frame rate. In normal operation the UVOT has a frame rate of 
90.6~s$^{-1}$ although modes with higher frame rates which read out only a
subset of the detector are available. For sources with count rates that are an
appreciable fraction of the frame rate, two
or more photons may arrive at a similar location on the detector within the same
CCD frame, and are counted as a single photon. This phenomenon is known as
coincidence loss \citep{fordham00b} and is analogous to pile-up in X-ray CCD
detectors. Source count rates are routinely corrected for coincidence loss in
the UVOT tasks in the {\sc heasoft} {\em Swift} {\sc ftools} software
package\footnote{{\sc heasoft} software can be found at:
  http://heasarc.gsfc.nasa.gov/docs/software/lheasoft/}, and details of the
corrections are described in \citet{poole08} and \citet{breeveld10}. The finite
number of frames in an exposure implies that the measured count rate follows a
binomial (rather than Poisson) distribution, and as the source count rate
approaches the frame rate of the CCD the photometric measurement errors become
larger rather than smaller \citep{kuin08}. 
The coincidence-loss correction has been calibrated up to 0.96
detected counts per frame (before correction for coincidence loss) 
in a 5~arcsec radius circular aperture \citep{poole08}, 
corresponding to a corrected count rate of 300~s$^{-1}$ 
in full-frame mode, beyond which
coincidence loss can no longer be accurately corrected, and this can be taken
to define the upper limit
to the dynamic range of UVOT.

Once per CCD frame the charge is transferred vertically to the read-out 
section. Sources add counts
to parts of the image that pass below them on the detector while the charge is
being transferred, so that bright sources give rise to vertical lines of 
enhanced brightness
on the image, aligned with the bright sources. These lines are commonly known
as read-out streaks\footnote{X-ray astronomers will note that the 
streaks of out-of-time events seen in X-ray CCD cameras are formed in an 
analogous process to the read-out streaks in 
UVOT.}; see Fig.~\ref{fig:schematic}. 

In this paper we describe how the read-out streaks can be used to make
photometric measurements with UVOT of stars which are too bright to be measured
through aperture photometry at the position of the star on the image.  While
the calibration obtained and verified 
in this paper is specific to the UVOT, 
the principles
apply equally to the XMM-OM. 

The paper is laid out as follows. Section~\ref{sec:instrument}
outlines the basic principles behind photometry using read-out
streaks. The catalogue of objects of known magnitudes which are used
to investigate photometry using read-out streaks and the method we
adopt to measure them are described in Section~\ref{sec:method}. The
results of our investigation are presented in Section
\ref{sec:results}. We discuss these results and develop a method to
measure photometry using read-out streaks in any UVOT pass-band and
window mode in Section~\ref{sec:discussion}. A demonstration of the
method is provided in the form of early-time photometry of the
exceptionally-bright gamma-ray burst GRB~080319B. Our conclusions are
presented in Section~\ref{sec:conclusions}. A step by step description
of our recommended procedure to derive photometry from read-out
streaks is given in Appendix A.

\section{Principles of photometry using read-out streaks in the UVOT}
\label{sec:instrument}

The UVOT detector employs an EEV CCD-02-06 frame-transfer CCD. At the
end of the integration period of each CCD frame, 
charge is transferred from the imaging area to the
frame-store in 290 steps, each taking 600~ns. The read-out streaks are
formed during this 174~$\mu$s interval; as only 256 of the 290 rows
correspond to the science imaging area, the effective exposure time of
a complete 256-row readout streak is 153.6~$\mu$s per frame.  For
comparison, in full-frame operation, the live exposure time
(i.e. excluding the frame transfer time) per 11.0329~ms frame is
10.8589~ms, so that the complete 256-row read-out streak would be
expected to contain 0.01415 times as many counts as the corresponding
object in the direct image. As the exposure is uniform at all points
along the readout streak the streak will have uniform brightness in
the vertical direction.
The speed at which the image is transferred
through the CCD is sufficiently large that coincidence loss should be
much less of a problem than in the direct image: the effective count
rate for coincidence loss is reduced by the ratio of counts in the
streak to the direct image divided by the extent of the streak in CCD
rows relative to the size of a photon splash. Thus if a photon splash
is considered to cover 3 pixels, this means that the effective count
rate for coincidence loss will be $0.01415\times 3 / 256 = 1.7\times
10^{-4}$ the count rate of the source in the direct image, implying a
$>9$~mag advantage in the bright limit for photometric
measurement with respect to the direct image. However, coincidence loss 
arising in a different part of the detector is expected to intervene well 
before such a large dynamic range is achieved. Charge is stripped from 
pores of the MCPs during the electron amplification, and the timescale
for the pores of the MCPs in the intensifier to recharge after
ejecting electrons is of order 0.3~ms. This dead-time in the MCP pores 
following an event may become the limiting factor by
giving rise to a second level of coincidence loss, which is
independent of the CCD \citep{fordham00a}.

\section{Method}
\label{sec:method}

\subsection{Input catalogue}
\label{sec:inputcat}

Investigation of the read-out streaks for photometry requires the
measurement of the read-out streaks of a sufficient sample of stars,
covering an appropriate magnitude range, with known magnitudes in a
photometric system which can be related systematically to that of the
UVOT. We decided to use stars from the Tycho-2 catalogue which fall
serendipitously within the fields of view of archival full-frame v-band UVOT
observations. To avoid issues related to image crowding, we selected
observations at Galactic latitudes of $b>20$ degrees so as to exclude
the Galactic Plane and Magellanic Clouds. The Tycho-2 catalogue
\citep{esa97} contains 2.5 million stars with photometry in two bands
from the sky mapper of the ESA {\em Hipparcos} satellite. The Tycho-2
V$_{T}$ passband covers a similar wavelength range to the UVOT v band,
and in combination with the Tycho-2 B$_{T}$ magnitudes a suitable
transformation to UVOT v band photometry can be made (see Section
\ref{sec:transformation}). The Tycho-2 catalogue contains stars with a
wide range of magnitudes with reliable photometry down to V$_{T}$=12
mag \citep{hog00}, which is approximately the limit to which
read-out streaks can be easily measured in UVOT images. 

To avoid problems with the transformation between Tycho-2 photometry
and UVOT v (see Section \ref{sec:transformation}), we excluded stars
of spectral type M from our analysis. Discrimination of M stars from
stars of earlier spectral type is poor using Tycho-2 B$_{T}-$V$_{T}$
colour, so we cross correlated our sample of Tycho-2 stars with the
Two Micron All Sky Survey \citep[2MASS;][]{skrutskie06}. Stars with
V$_{T}-$J$>3.0$ are likely to be M stars and were excluded from our
sample. After these stars were excluded, we measured the read-out streaks
 for a total of 160 stars.

\subsection{Measurement of read-out streaks}
\label{sec:measurement}

\begin{figure}
\includegraphics[width=84mm]{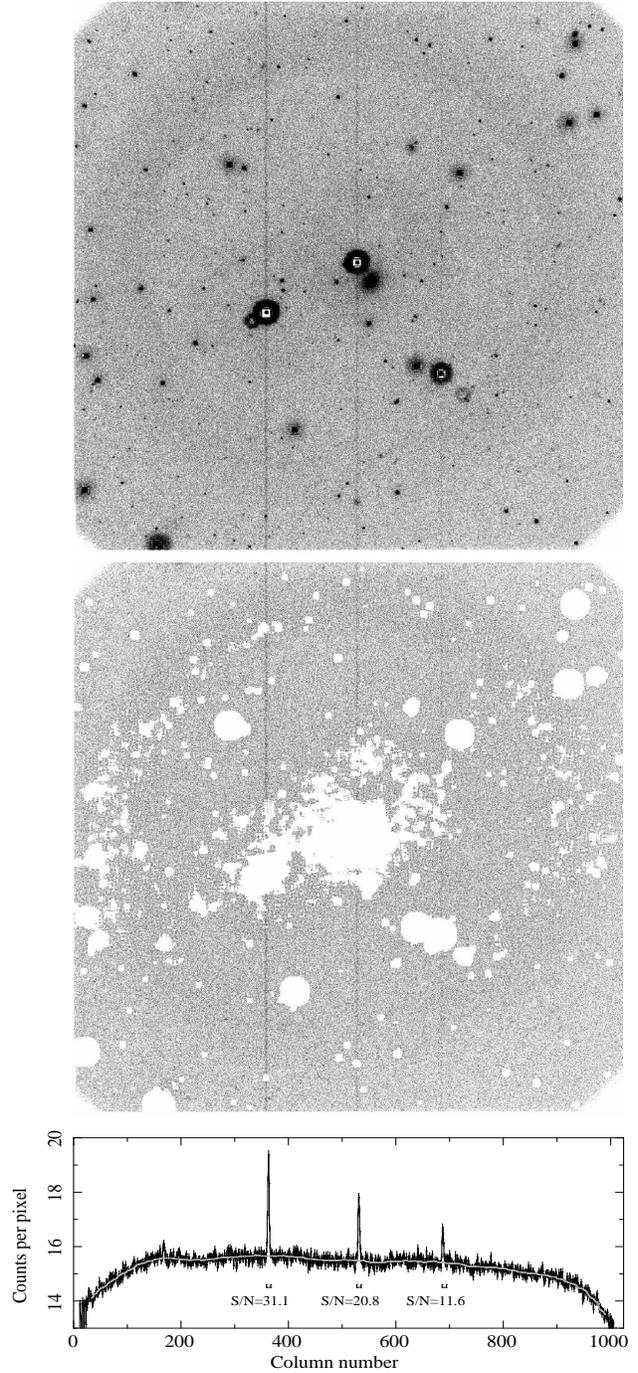}
\caption{Illustration of the measurement process for read-out streaks. Top
  panel: raw v band image from observation number 00030875001. The image was
  binned onboard the spacecraft by a factor of 2 in x and y, as is common for
  UVOT images. A logarithmic
  colour table has been employed to emphasize faint features including the 
read-out streaks. Middle panel: the same image after bright stars and features
have been masked, as described in Section~\ref{sec:measurement}. Bottom panel:
the mean pixel brightness in each column of the masked image. The grey line
shows the 128-column median which is used for background estimation. 
The read-out
streaks from the 3 brightest stars are significantly detected. The apertures
used to measure the brightness of the streaks are shown below them, and labelled
with the signal to noise ratios.
}
\label{fig:schematic}
\end{figure}

Read-out streaks are precisely vertical in raw UVOT images, but this
is not the case after images have been corrected for distortion
(primarily from the fibre taper connecting the phosphor screen to the
CCD) and rotated so that x and y coordinates correspond to equatorial
coordinate axes. Therefore read-out streaks are best measured from the
raw UVOT images, after correction for modulo-8 noise, but before the
distortion corrections.  In practice the read-out streak can never be
measured over the full vertical extent of the image because the direct
image of the star responsible for the read-out streak, and any other
stars which are located in the same column, will be superimposed on
it. Therefore the brightness of the streak must be measured from parts
of the streak which are not contaminated by objects in the image. 

The brightness of the read-out streaks was measured according to the
following procedure. Note that in the description that follows, the
terms pixel, rows and columns refer to unbinned 
UVOT image pixels, rows and columns of
image pixels respectively. First, bright sources with count rates
exceeding 40 count~s$^{-1}$ were identified in the image and masked
with circular regions of radius 24~arcsec. This separate step is
needed for bright stars because they are surrounded by dark regions
which result from coincidence loss, which would not be identified in a
regular source searching process. Next, each column of the image was
searched for enhanced-brightness pixels corresponding to sources in
two steps. First, individual pixels exceeding the median pixel value of
the column by more than 3$\sigma$, or by more than 3 counts if the
median of the column is $\le 1$, are flagged. In the second step, the
column is smoothed with a 10-pixel boxcar filter to improve
sensitivity to faint sources before again flagging pixels which exceed
the median of the column by more than 3$\sigma$, or by more than 3
counts if the median of the column is $\le 1$. Pixels flagged in
either step are then masked from the column. The rationale for
performing this operation column by column rather than simply
source-searching the image is that it prevents the read-out streaks
themselves from being erroneously identified as sources. Next, the
columns are collapsed to a single row containing, for each column, the
mean value of all non-masked pixels in the column. Read-out streaks
are then identified using a 16-column sliding box along the row, at a
6-sigma threshold with respect to the median counts measured in a
128-column sliding box. The count rate in the 16-pixel aperture is
then background-subtracted using the 128-column median, and scaled to a
16-row section of the image (i.e. multiplied by a factor of 16
assuming an unbinned image).  The scaling in this stage is arbitrary,
but $16$~pixels~$\times$~$16$~pixels is a convenient equivalent-aperture in
which to measure the count rate of the read-out streak because it has
a similar sky area to the 5~arcsec radius circular aperture normally used 
for photometry of point sources in UVOT 
images\footnote{The sky area of the aperture used for measuring 
the read-out streaks is slightly smaller than that used for aperture 
photometry of point sources in the image, but this difference is 
offset by the lossless 
redistribution of photons in the vertical direction in the read-out streak: 
photons in the wings of the point-spread function are not lost from the 
read-out streak when they are displaced in the vertical direction.},
as well as corresponding to precisely $2\times 2$ CCD pixels (which
are subsampled by a factor of 8 in the event centroiding process, as
described in Section \ref{sec:introduction}). The size of the 128-pixel 
sliding box used for background is chosen as a compromise between tracing 
the variation in background over the image (see Fig.~\ref{fig:schematic}) 
and minimising the contribution of the background estimation to the 
statistical uncertainty of the read-out streak measurement.
Finally, the count rate
of the read-out streak is corrected for the large-scale sensitivity
variations (evaluated at the position of the star in the image)
and time-dependent sensitivity degradation of the UVOT
\citep{breeveld10}.

\subsection{Transformation from the Tycho-2 to UVOT photometric system}
\label{sec:transformation}

\begin{figure}
\includegraphics[width=64mm,angle=270]{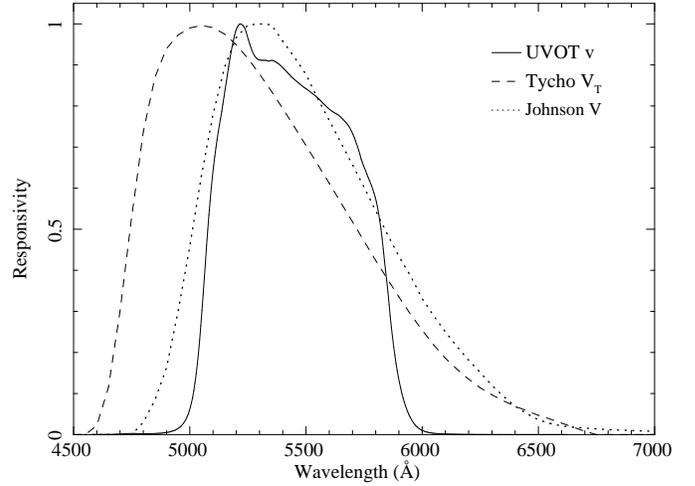}
\caption{Responsivity curves for the UVOT v (solid line), 
Tycho-2 V$_{T}$ (dashed line) and Johnson V (dotted line) passbands.}
\label{fig:filters}
\end{figure}

\begin{figure}
\includegraphics[width=64mm,angle=270]{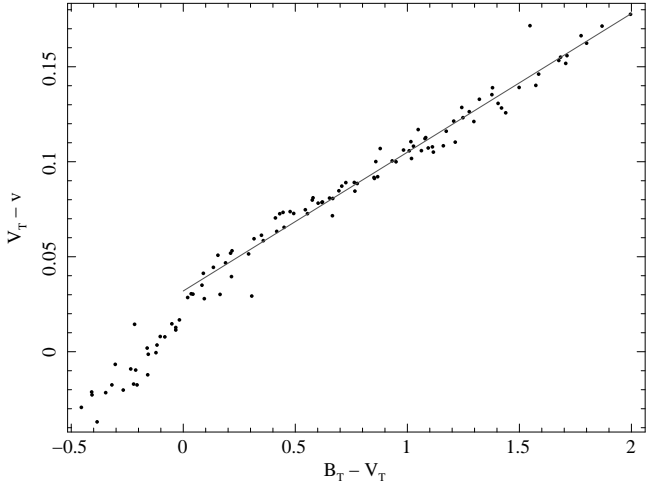}
\caption{Difference between UVOT v and Tycho-2 V$_{T}$ magnitudes as a function
  of Tycho-2 B$_{T} - $V$_{T}$ colour. The points correspond to synthetic
  photometry of stars from the
  \citep{pickles98} spectral atlas. Stars of spectral type M are not shown. The
  grey line shows the fit to the data from which the colour transformation
  given in Section \ref{sec:transformation} is derived. 
 }
\label{fig:transform}
\end{figure}

Responsivity curves for the UVOT v band, Johnson V band, and Tycho-2
V$_{T}$ band are shown in Fig.~\ref{fig:filters}. While all three cover
approximately the same wavelength range, there are significant
differences in the shapes of the curves, so that translation between
the different photometric systems is colour-dependent. To derive a
transformation between Tycho-2 V$_{T}$ and UVOT v magnitudes, we
generated synthetic B$_{T}-$V$_{T}$ and v$-$V$_{T}$ colours for the
stars of the \citet{pickles98} spectral
library. Fig. \ref{fig:transform} shows the resulting colour-colour
distribution for spectral types K and earlier. For stars with
B$_{T}-$V$_{T} > 0$ there is a tight linear relation between
B$_{T}-$V$_{T}$ and v$-$V$_{T}$. From a least-squares fit to the data
points with B$_{T}-$V$_{T} > 0$ (shown as a grey line in
Fig. \ref{fig:transform}) we obtain the following transformation.
\begin{equation}
v = {\rm V}_{T} - 0.032 -0.073 ({\rm B}_{T}-{\rm V}_{T})
\end{equation}
The rms scatter of the datapoints about this relation is 0.006 mag
for B$_{T}-$V$_{T} > 0$. Stars with B$_{T}-$V$_{T} < 0$ deviate from
this linear relation, but this is not of concern because all of the
Tycho-2 stars used in our read-out streak analysis have B$_{T}-$V$_{T}
> 0$.

Stars of spectral type M were not used to derive the colour
transformation because they show a much larger scatter in the
B$_{T}-$V$_{T}$ and v$-$V$_{T}$ relation than stars of earlier
spectral type; a similar problem besets the transformation of 
Tycho-2 magnitudes to Johnson V \citep{esa97}.

\section{Results}
\label{sec:results}

\begin{figure}
\includegraphics[width=84mm]{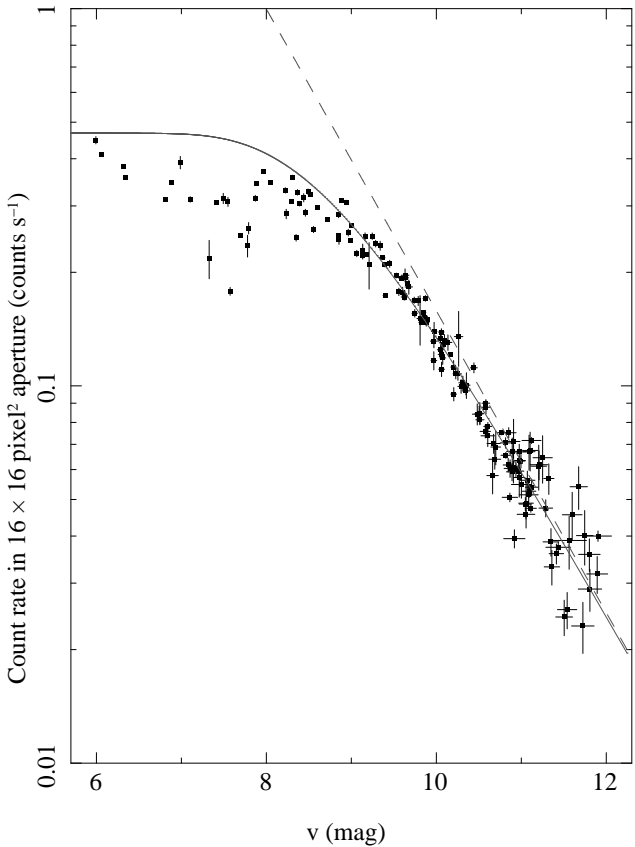}
\caption{Count rates of the read-out streaks originating from Tycho-2 stars
  measured in a $16$~pixel~$\times$~$16$~pixel aperture. The grey dashed line 
  corresponds
  to the predicted relation given in Section \ref{sec:instrument} based on the
  relative exposure times of the read-out streak and the static image; it is
  not a fit to the data. The grey solid line shows the expected relation in the
  presence of coincidence loss as described by Equation~\ref{eq:coiloss} with
  $t_{MCP}=2.36\times 10^{-4}$~s.
 }
\label{fig:vrate}
\end{figure}

Fig. \ref{fig:vrate} shows the relation between v magnitude derived from 
Tycho-2 and the 
count rate of the read-out strip in a $16$~pixel~$\times$~$16$~pixel aperture
for the sample of Tycho-2 stars observed in full-frame UVOT v-band images. 
As described in
Section~\ref{sec:instrument}, the fixed ratio of exposure time in the
read-out streak and the static image implies that the readout streak
will receive 0.01415 times as many counts as the direct image. This
ratio must be divided by a factor of 128 to scale to a 16 pixel
section of the readout streak (corresponding to our 
$16$~pixel~$\times$~$16$~pixel aperture), 
so that the zeropoint appropriate to convert
read-out-streak count rate to magnitude in such an aperture will be
9.89 magnitudes brighter than the corresponding zeropoint used for
aperture photometry of the static image; for the v band of UVOT this
implies a zeropoint of 8.00 for photometry using the read-out
streak. The dashed line in Fig.~\ref{fig:vrate} shows the count-rate
to magnitude relation predicted from this zeropoint. While the
measured count-rates are clustered around the predicted relation at
the faintest magnitudes (v$>11$), they are systematically below it at
brighter magnitudes. For v$<8$ the count-rates show a large scatter
and show little correlation with the magnitude of the source. As
described in Section~\ref{sec:instrument}, these magnitudes are not
bright enough for the diminished count-rates to be due to coincidence
loss on the CCD during frame transfer. A more likely culprit is the
coincidence loss of photons which arrive while the MCP pores are
recharging, predicted by \citet{fordham00b}. As this coincidence loss
arises in the MCPs rather than the CCD, it is independent of the CCD
frame time. We can assume that by the third in-series MCP in the
intensifier, the area of the MCP which is utilised in the detection of
incoming photons from a point source is large enough that the process
can be treated as coincidence loss in a single-pixel
detector. Starting from equation 4 of \citet{fordham00b}
\begin{equation}
C_{o} = 1 - e^{-C_{i}}
\end{equation}
where $C_{o}$ is the mean number of counts observed per frame (i.e. after
coincidence loss) and $C_{i}$ is
the mean number of incoming counts per frame (i.e. before coincidence loss).
In this case a frame is the timescale required for the MCP to recharge, which
we will call $t_{MCP}$, and all counts from the source contribute to
coincidence loss, not just those which arrive during the formation of the
readout streak. 
If the ratio of the integration time of the static image to 
the frame-transfer time is $S$, the count rate of a source in the static image 
will be  
 $S$ times the count rate that will be measured in a 
16~pixel section of the 
read-out streak. 
Hence the numbers of
counts per frame are related to the count rates before and after correction for
coincidence loss as 
\[
C_{i} = S\ R_{i}\ t_{MCP} 
\]
and
\[
C_{o} = S\ R_{o}\ t_{MCP}   
\]
respectively, where $R_{i}$ is the incoming count rate in a 
$16$~pixel~$\times$~$16$~pixel
section of the read-out streak before coincidence loss and
$R_{o}$ is the observed count-rate with coincidence loss. Substituting and 
rearranging, we
obtain:
\begin{equation}
R_{o} = \frac{1 - e^{-(S\ R_{i}\ t_{MCP})}}{S\ t_{MCP}} 
\label{eq:coiloss}
\end{equation}
The values of $S$ corresponding to the full-frame and windowed operating modes 
of UVOT are 
given in Table~\ref{tab:windowmodes}.

\begin{table}
  \caption{Characteristics of the full-frame and windowed operating modes of 
UVOT that are important for read-out streak photometry. $S$ is the ratio 
of the static image exposure time to the frame-transfer exposure time 
(see Section \ref{sec:results}). Max count rate refers to the maximum 
coincidence-loss-corrected count rate of a read-out streak 
in a 16~pixel~$\times$~16~pixel aperture 
for which we recommend using read-out streak photometry.}
\label{tab:windowmodes}
\begin{tabular}{lccc}
Window mode&CCD Frame time&$S$&Max count rate\\
           &(ms)&&(s$^{-1}$)\\
\hline
&&&\\
Full frame&11.0329&9049&0.30\\
Large window&5.417&4369&0.62\\
Small window&3.600&2855&0.95\\
\hline
\end{tabular}
\end{table}

As the timescale for MCP recharge $t_{MCP}$ is not known a priori, we
performed a $\chi^{2}$ fit to the data in Fig.~\ref{fig:vrate}. In the
fitting, magnitude uncertainties on the Tycho-2 stars were translated
into count-rate uncertainties and added in quadrature to the
measurement errors on the count rates. Stars brighter than v=9, for
which the read-out streaks exhibit an increasing degree of scatter
with brightness, were excluded from the fit.  The fitting yields a
best-fit $t_{MCP}=2.36\pm0.03\ \times 10^{-4}$~s. The relation for
count-rate against magnitude corresponding to $t_{MCP}=2.36\times
10^{-4}$~s is shown as the solid curve in Fig.~\ref{fig:vrate}. It
reproduces the overall trend of the data well for sources fainter than
9th magnitude and the apparent count-rate ceiling at 
$\sim$0.45~s$^{-1}$. However, coincidence loss does not explain the large
scatter in count rates seen for sources with v$<8$. There are (at
least) two possible causes for this large scatter at bright
magnitudes. It might be due to the electron wells of the CCD becoming
over-full, leading to charge bleeding into the relevant CCD column so
that individual photon splash events can no longer be identified as
the frames are read out; the degree to which counts are lost might
depend on the positioning of the source in the field of view, and/or
with respect to the boundaries of the CCD columns. Alternatively, it
could be due to variations of the electron mobility with position on
the MCPs, perhaps as a result of local ageing of the MCPs (which would
imply a dependence on time as well as position).  Whatever the cause,
no useful information on source brightness can be gleaned from the
read-out streaks for sources brighter than magnitude 8 in v.

\begin{figure}
\includegraphics[width=84mm]{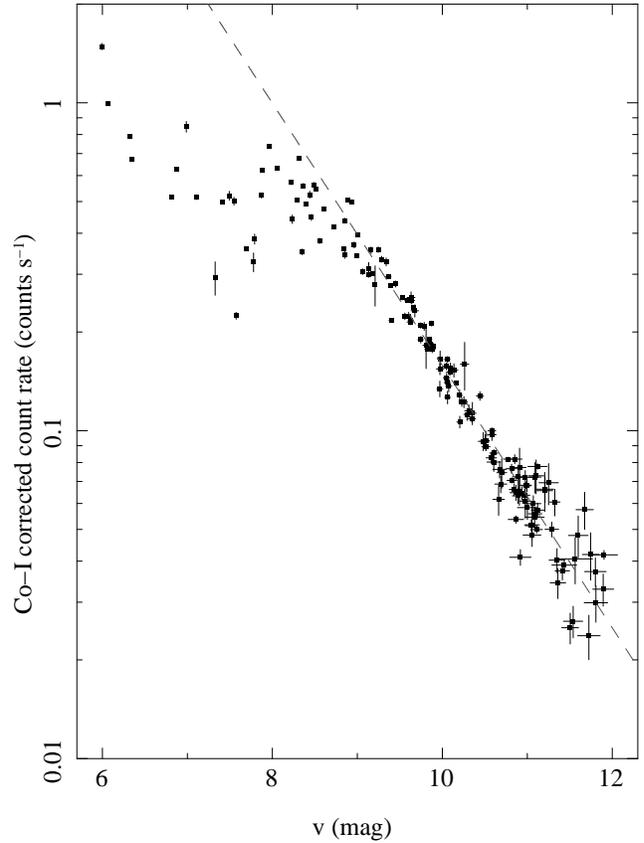}
\caption{Count rates of the read-out streaks originating from Tycho-2
  stars measured in a $16$~pixel~$\times 16$~pixel aperture after
  correction for coincidence loss via Equation~\ref{eq:coicorr} and
  $t_{MCP}$=0.236~ms. As in Fig.~\ref{fig:vrate}, the dashed line
  corresponds to the predicted relation given in Section
  \ref{sec:instrument} based on the relative exposure times of the
  read-out streak and the static image.
}
\label{fig:vrate_coi}
\end{figure}

\begin{figure}
\includegraphics[width=84mm]{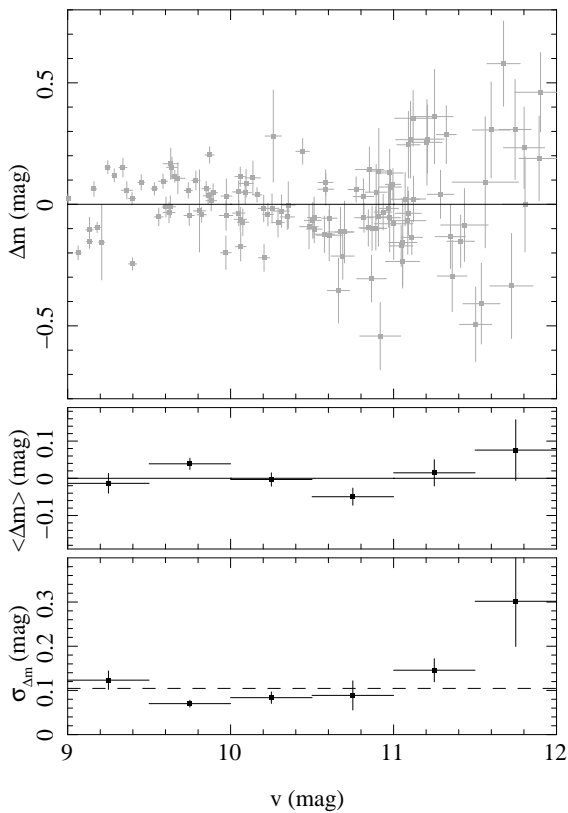}
\caption{Top panel: differences $\Delta m$ between the v magnitudes
  obtained from Tycho-2 and the v magnitudes obtained from the
  coincidence-loss corrected count-rates of the UVOT read-out
  streaks. Uncertainties are the quadrature sums of the errors on the
  count-rates and the Tycho-2 magnitudes. Middle and bottom panels:
  mean and dispersion of $\Delta m$ respectively in 0.5 magnitude
  bins. The dashed line in the bottom panel shows the best-fit
  dispersion over the full $9<$v$<12$ magnitude interval shown.  
}
\label{fig:deltam}
\end{figure}

Rearranging equation~\ref{eq:coiloss} we obtain:
\begin{equation}
R_{i} = -\frac{\log_{e}(1.0 - (S\ R_{o}\ t_{MCP}))}{S\ t_{MCP}}
\label{eq:coicorr}
\end{equation}
which can be used together with our best-fit $t_{MCP}=2.36\times
10^{-4}$~s to correct the measured count-rate in the read-out streak
for coincidence loss. Fig.~\ref{fig:vrate_coi} shows the corrected
count-rates for the the Tycho-2 stars against their v magnitudes. As
in Fig.~\ref{fig:vrate}, the dashed line gives the expected relation
between magnitude and count-rate. The coincidence-loss-corrected
count-rates trace the dashed line well for v$>9$ mag. We now look in
more detail at the photometric precision which might be achieved using
read-out streak measurements. The top panel of Fig.~\ref{fig:deltam} shows the
differences $\Delta m$ between the v magnitudes derived from the
Tycho-2 magnitudes and the v magnitudes obtained from the
coincidence-loss-corrected read-out streak measurements as a function
of v magnitude (from Tycho-2) of the individual stars. 
The uncertainties on $\Delta m$ are the
sums in quadrature of the Tycho-2 and UVOT uncertainties. 

To look for any systematic trends of photometry with magnitude and to
estimate the level of photometric precision that can be achieved
through measurements of the read-out streaks, we have used the
maximum-likelihood approach outlined by \cite{maccacaro88} to measure
the mean and intrinsic standard deviation of the $\Delta m$
distribution (which we assume to be Gaussian), taking into account the
measurement errors, in half-magnitude bins. The means are shown in the
middle panel of Fig.~\ref{fig:deltam} and do not show any systematic
trend with magnitude. Except for the faintest magnitude bin, where the
error is larger, the mean values of $\Delta m$ ($\langle \Delta
m \rangle$) are within $\pm0.05$ mag of $\langle \Delta m\rangle =0$; the
faintest magnitude bin is consistent with $\langle \Delta m
\rangle=0$ within its uncertainty.  Over the full $9<$v$<12$
magnitude range $\langle \Delta m
\rangle=-0.005\pm0.005$.  The standard deviations
$\sigma_{\Delta m}$ are shown in the bottom panel. Apart from the
faintest magnitude bin (which again has a larger uncertainty than the
others) the best-fit values of $\sigma_{\Delta m}$ are between 0.07
and 0.13 mag. Over the full magnitude range, the best-fit
$\sigma_{\Delta m}=0.104\pm0.004$ mag, and this is shown in the bottom
panel of Fig.~\ref{fig:deltam} as a dashed line. This $\sigma_{\Delta m}$ 
represents the systematic error on the read-out streak photometry which 
remains after the statistical uncertainty is accounted for.

\section{Discussion}
\label{sec:discussion}

\subsection{Using read-out streaks for photometry}

\begin{table}
\caption{Zeropoints for read-out streak photometry in full-frame 
or windowed mode, 
for a 16~pixel~$\times$~16~pixel aperture, in the 7 UVOT imaging passbands. 
The zeropoints are in the UVOT Vega magnitude system; for AB$-$Vega 
magnitude offsets see \citet{breeveld11}.}
\label{tab:zeropoints}
\begin{tabular}{lccc}
Filter&Full-frame&Large-window&Small-window\\
      &zeropoint&zeropoint&zeropoint\\
\hline
&&&\\
UVW2&7.49&8.28&8.74\\
UVM2&6.96&7.75&8.21\\
UVW1&7.55&8.34&8.80\\
u&8.45&9.24&9.70\\
b&9.22&10.01&10.47\\
v&8.00&8.79&9.25\\
White&10.40&11.19&11.65\\
\hline
\end{tabular}
\end{table}

\begin{figure}
\includegraphics[width=84mm]{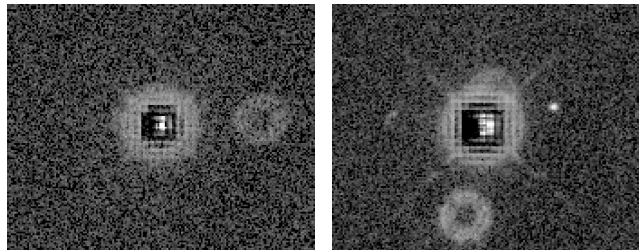}
\caption{Identifying heavily saturated stars. The star on the left has a v
  magnitude of 9.3 and is suitable for read-out streak photometry, while the
  star on the right is of v magnitude 7.8, and is too bright for read-out
  streak photometry. Note the presence of diffraction spikes, the larger
  extent of the modulo-8 distorted region, and the crescent like appearence of
  the core in the star on the right. 
}
\label{fig:saturated}
\end{figure}

The use of UVOT read-out streaks for photometry relies on the
calibration or knowledge of four instrumental properties: the
effective area as a function of wavelength, the zeropoint for direct
imaging, the ratio of exposure times between the read-out streak and
the static image in the CCD, and the effective recharge timescale for
the MCP intensifier. The effective area curves and zeropoints have
been calibrated to high precision for the 7 UVOT imaging bands. The
ratio of exposure times is fixed by the operation of the CCD and the
effective recharge timescale for the MCPs has been determined in
Section~\ref{sec:results} ($t_{MCP}=$0.236~ms); both of these
quantities are independent of the imaging passband in question.  Thus
all the required elements are in place to enable read-out streaks to
be used for photometric measurements in all the UVOT imaging
passbands.  Table~\ref{tab:zeropoints} gives the zeropoints for
read-out streak photometry, determined from the standard imaging
zeropoints \citep{breeveld11} and the ratio of the CCD integration and
frame-transfer times.  In Section~\ref{sec:results} we showed that the
read-out streaks produce good v-band photometry for sources from
  the detection limit of v=12 up to v=9, but in general one will not
know the magnitude of the source a priori, so the bright-source
threshold must instead be specified in terms of observed count rate in
the read-out streak. Inspection of Fig.~\ref{fig:vrate_coi} suggests
that coincidence-loss-corrected count rates up to 0.3~s$^{-1}$ will
produce robust photometry for full-frame imaging\footnote{Note that
  this cut-off in count rate, i.e. a horizontal line in
  Fig.~\ref{fig:vrate_coi}, translates via Table~\ref{tab:zeropoints}
  to a nominal bright-limit of v=9.31, which is more conservative than
  the bright limit of v=9.0, discussed so far, 
  for a v magnitude which is known a-priori.}; equivalent maximum count
rates for the windowed modes are provided in Table
\ref{tab:windowmodes}. This bright-source limit is 2.4 magnitudes
brighter than the coincidence-loss limit for standard aperture
photometry of a point source in full-frame imaging.  We note that the
large scatter in the read-out streak brightness for v$<8$ means that
some much brighter stars stray into the regime of count rates
$<0.3$~s$^{-1}$. In practice, these very heavily saturated stars can
be discriminated by their appearance, as illustrated in
Fig.~\ref{fig:saturated}.

We outline our recommended procedure for obtaining photometry from 
read-out streaks in Appendix A and provide a brief discussion of the 
potential extension of the method to XMM-OM in Appendix B.

\subsection{Photometric scatter}
\label{sec:scatter}

In the measurement of the read-out streaks outlined in
Section~\ref{sec:measurement}, we estimated the uncertainties assuming
Poisson statistics, purely on the measured number of counts within the
read-out streak aperture. The uncertainty on the background was
assumed to be negligible, as it was determined over a much larger
number of columns than the measurement of the read-out streak. The
Poisson errors were scaled by the same factor as the count rates in
the coincidence-loss correction stage
(Equation~\ref{eq:coicorr}). However, in Section~\ref{sec:results} we
found an additional systematic component to the photometric scatter between
the Tycho-2 magnitudes and the UVOT read-out-streak magnitudes of
around 0.1 mag after taking into consideration the uncertainties on
the Tycho-2 magnitudes and the Poisson uncertainties on the read-out
streak measurements. We can identify at least 2 sources of uncertainty which
 contribute to this additional scatter, and a third source that may. 
The first is inherent to
the background determination. Usually one might assume that the
background is intrinsically smooth on 16 pixel (8 arcsec) scales, and this
would be a reasonable assumption for the genuine zodiacal background in the
UVOT images. However, the column-by-column background also contains the
read-out streaks of all the sources in the image, which will contribute genuine
scatter on the scale of the 16 pixel aperture we have used to measure the
read-out streaks. Secondly, the coincidence loss due to the MCPs changes the
inherent uncertainties on the count rates in the read-out streaks from 
Poisson errors to binomial
errors \citep{kuin08}, which are larger. Unfortunately, unlike the case
considered by \citet{kuin08},  the
read-out streaks are superimposed on a significant background which is
accumulated over a different time interval 
(during the integration of the static image)
and for which MCP-related coincidence loss is negligible. Thus in the case of
the measured read-out streaks, the uncertainties are neither pure Poisson nor
pure binomial. Our recommended maximum count rate of 0.3~s$^{-1}$ bright limit
for read-out-streak photometry corresponds to 0.5 incident counts per MCP
frame; comparison with Fig.~1 of \citet{kuin08}, shows that this limit 
corresponds to a
binomial uncertainty which is 1.5 times as large as the Poisson error, so this
is a significant effect. These two effects play off against each other:
uncertainties in the background will produce the most significant photometric
errors for faint read-out streaks, while the non-Poissonian nature of the
errors induce photometric scatter primarily for bright read-out streaks. 
The third, potential, effect is that we do not have a measure of the uniformity
of the MCP recharge time over the detector or its long-term evolution as the
MCPs age. Any such variations in the recharge time of the MCPs will correspond
directly to a scatter in the value of $t_{MCP}$ which should be employed in the
coincidence-loss correction, which will in turn induce photometric scatter in 
data
which is corrected for coincidence loss assuming $t_{MCP}=0.236$~ms. Such an
effect will 
produce photometric scatter which is largest for the brightest 
read-out streaks. 

We do not have a theoretical prescription for any of these additional
sources of photometric uncertainty, so we resort to the empirical
estimate of the photometric scatter which is seen in addition to the
simple Poisson errors. In keeping with the results presented in
Fig.~\ref{fig:deltam} and Section~\ref{sec:results}, we suggest that a
systematic error of 0.1 mag is added in quadrature to all magnitudes
derived from read-out streaks. 

\subsection{Verification in b}

To verify the read-out streak photometry procedure and the photometric accuracy derived above, in a different UVOT band, we move to the UVOT b band and once again make use of the Tycho-2 catalogue. The UVOT b band extends about 400\AA\ to the red of the Tycho-2 B$_{T}$ band, so the transformation from B$_{T}$ to b magnitude has a large colour term in B$_{T}-$V$_{T}$. Making use of the \citet{pickles98} spectral library, we obtain the following transformation for stars with B$_{T}-$V$_{T} > 0.4$, excluding M stars.
\begin{equation}
b = {\rm B}_{T} + 0.036 -0.270 ({\rm B}_{T}-{\rm V}_{T})
\label{eq:btransform}
\end{equation}
The rms scatter of the datapoints about this relation is 0.009 mag. 
Stars with B$_{T}-$V$_{T} < 0.4$ do not follow this relation. 

We then measured the read-out streaks in the b-band,
following the procedure outlined in Appendix A,  for a subset of
the Tycho-2 stars described in Section \ref{sec:inputcat}, which have
B$_{T}-$V$_{T} > 0.4$ and which have b-band observations with UVOT. This sample
consists of 74 stars. 

\begin{figure}
\includegraphics[width=84mm]{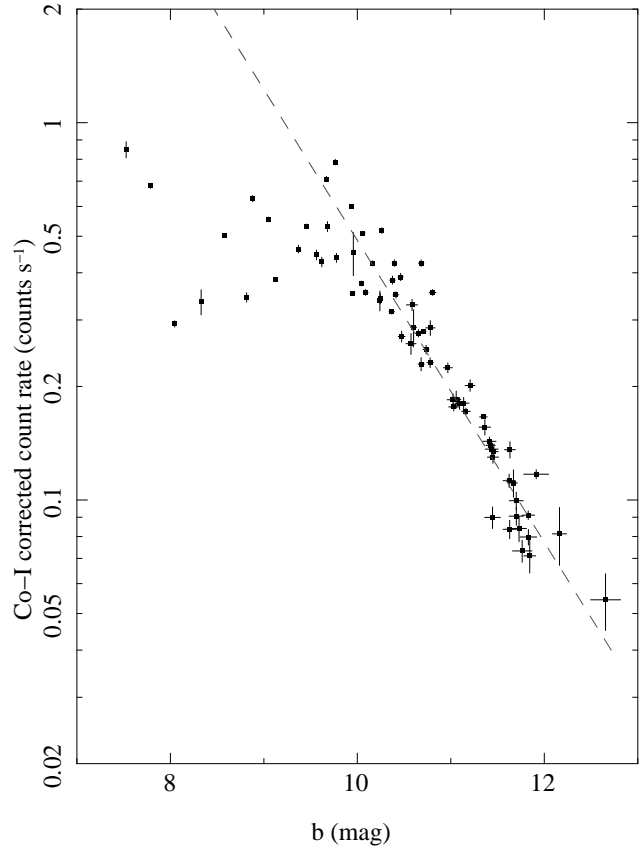}
\caption{Count rates of the read-out streaks from Tycho-2
  stars measured in the UVOT b band after
  correction for coincidence loss via Equation~\ref{eq:coicorr} and
  $t_{MCP}$=0.236~ms. The dashed line
  corresponds to the predicted relation given in Section
  \ref{sec:instrument} based on the relative exposure times of the
  read-out streak and the static image.
}
\label{fig:brate_coi}
\end{figure}

The coincidence-loss-corrected count-rates are shown in 
Fig.~\ref{fig:brate_coi} as a function of UVOT b magnitude, as derived from
the Tycho-2 photometry. The data points follow closely the behaviour
seen in the v band: they follow the relationship predicted from the
ratio of the exposure times of the static image and the read-out
streak for count rates below 0.3~s$^{-1}$, and deviate from
this relationship increasingly at brighter magnitudes. 

\begin{figure}
\includegraphics[width=84mm]{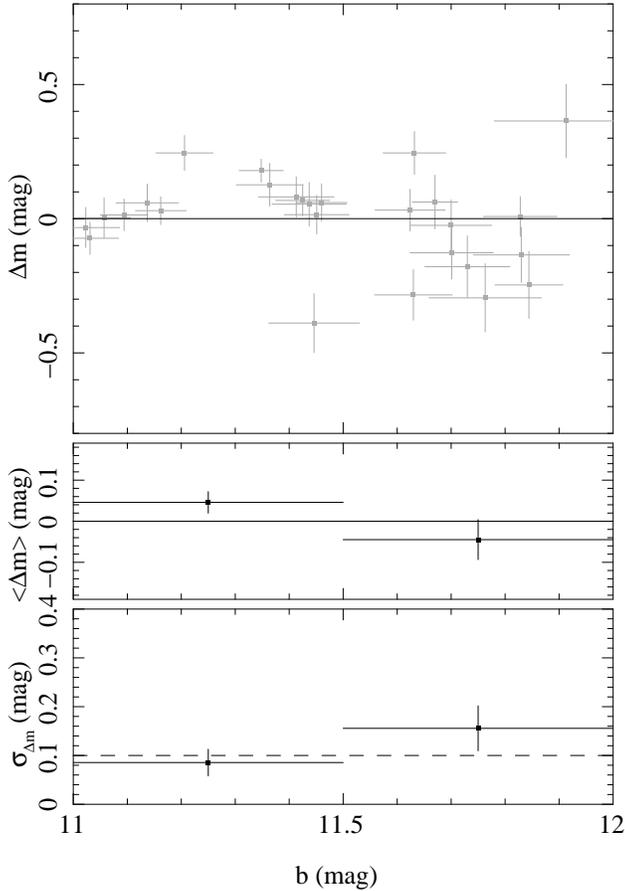}
\caption{Top panel: differences $\Delta m$ between the b magnitudes
  obtained from Tycho-2 and the b magnitudes obtained from the
  coincidence-loss corrected count-rates of the UVOT read-out
  streaks. Uncertainties are the quadrature sums of the errors on the
  count-rates and the Tycho-2 magnitudes. Middle and bottom panels:
  mean and dispersion of $\Delta m$ respectively in 0.5 magnitude
  bins. The dashed line in the bottom panel shows the 0.1 magnitude
  systematic advocated in Section \ref{sec:scatter}.  
}
\label{fig:bdeltam}
\end{figure}

The throughput of UVOT is higher in b than in v, but the Tycho-2
photometry is less precise in B$_{T}$ than in V$_{T}$ \citep{hog00},
with the consequence that the dynamic range over which UVOT b-band read-out
streak measurements can be compared to magnitudes derived from
Tycho-2 is considerably smaller than for the v band. Figure
\ref{fig:bdeltam} shows the differences between the b magnitudes
derived from read-out streaks and the b magnitudes derived from
Tycho-2 photometry, in the magnitude range for which the count rates
are below 0.3~s$^{-1}$. Note that in this figure the 0.1
magnitude systematic error (see Section \ref{sec:scatter}) has not
been included in the uncertainties so that the magnitude of this
systematic can be verified. As can be seen in this figure, there is no
systematic offset between the predicted and observed read-out-streak
magnitudes, and the systematic component of the uncertainty is
consistent with the 0.1 magnitudes derived in Section
\ref{sec:scatter}.

\subsection{Verification in UVM2}

Although there is no comparable source of photometry to the Tycho-2
catalogue for the UVOT bands other than v and b, it is possible to verify
that the read-out streaks produce viable photometry in the
ultraviolet, for objects for which photometry can be synthesized using
archival IUE spectra. The UVM2 band is the best UV band to perform
this test because the response of UVM2 is almost completely contained
within the IUE spectral range. In contrast UVW1 and UVW2 have red
wings to their responses which extend beyond the long wavelength limit
of IUE \citep[see Figure 2 of ][]{breeveld11}.  
The O stars HD\,15570
and HD\,168076 have suitable IUE spectra and have been observed in the
UVM2 band with UVOT. Details of these two targets are provided in
Table~\ref{tab:uvm2stars}. The IUE large-aperture, low-resolution
NEWSIPS spectra were obtained from the Mikulski Archive for Space
Telescopes (MAST) and were combined to form a single spectrum per
object. The spectra were then convolved with the UVM2 response to
obtain synthetic magnitudes; the uncertainties on these magnitudes are
dominated by systematics in the IUE spectra, which we assume to be 10
per cent \citep{massa00}. The read-out streaks from these two stars
were measured using the procedures outlined above (including the 0.1
mag systematic error), and compared to the synthetic photometry from
IUE. As seen in Table~\ref{tab:uvm2stars}, the UVM2 magnitudes
measured from the read-out streaks show good consistency with the
synthetic magnitudes measured from IUE.

\begin{table}
\caption{Comparison of UVM2 magnitudes obtained from read-out streaks with synthetic magnitudes from IUE.}
\label{tab:uvm2stars}
\begin{tabular}{lccc}
Star&Swift&UVM2 from&UVM2 from\\
&observation&IUE (mag)&UVOT read-out\\
&number&&streak (mag)\\
\hline
&&&\\
HD\,168076&00044156001&$8.91\pm0.10$&$8.73\pm0.10$\\
HD\,15570&00090050001&$10.45\pm0.10$&$10.44\pm0.10$\\
\hline
\end{tabular}
\end{table}

\subsection{Demonstration: the naked eye GRB 080319B}

\begin{figure}
\includegraphics[width=84mm]{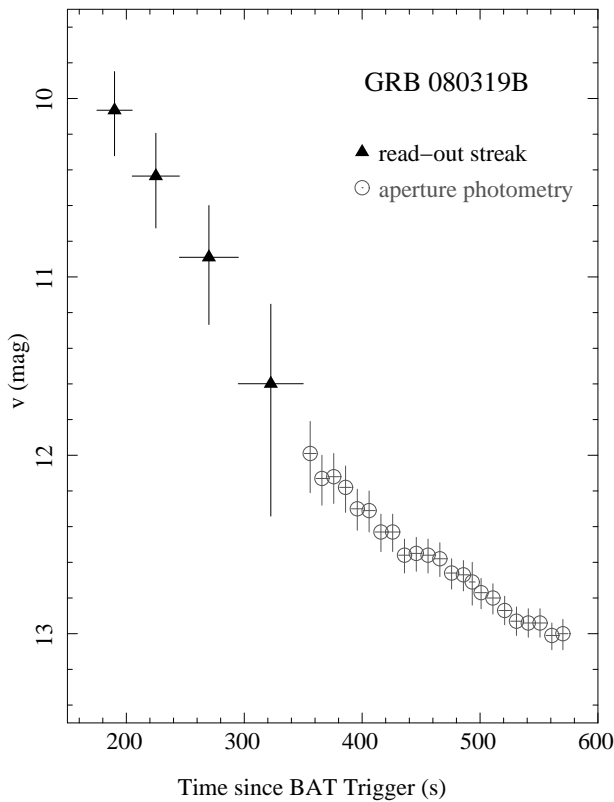}
\caption{Early-time UVOT v band photometry of GRB 080319B. The triangles are
  measurements obtained from read-out streaks while the source is too bright to
  be measured from the direct image, and circles show the aperture
  photometry at later times from \citep{racusin08}.
}
\label{fig:grb080319b}
\end{figure}

GRB\,080319B had the brightest optical afterglow of any GRB so far observed
with {\em Swift}. The afterglow was sufficiently bright that the v-band
event-mode UVOT data taken less than 350~s after the BAT trigger could not be
used due to coincidence loss \citep{racusin08}. Here we show that the read-out
streaks can be used to obtain photometry between 170~s and 350~s after the
trigger (i.e. from the beginning of the v finding-chart exposure until the
afterglow is faint enough that normal aperture photometry can be used). The
event mode data between 170~s and 350~s was split into four time intervals, and
in each time interval an image in raw coordinates was formed. In each image the
read-out streak was measured following the procedure described in Appendix
A. The photometry so derived is listed in Table~\ref{tab:grb080319b}, and shown
together with the aperture photometry for the remainder of the finding-chart
exposure, taken from the supplementary information of \citet{racusin08}, in
Fig.~\ref{fig:grb080319b}.

\begin{table}
\caption{Early-time UVOT v band photometry of GRB 080319B measured from
  read-out streaks in images formed in raw coordinates from event-mode data. 
The first column, $T-T_{0}$,
  gives the mid-time of the image after the BAT trigger, and the second column
  gives the duration of the image. }
\label{tab:grb080319b}
\begin{tabular}{lcc}
$T-T_{0}$ (s)&Duration (s)&v (mag)\\
\hline
&&\\
190.0&30.0&$10.07^{+0.26}_{-0.22}$\\
225.0&40.0&$10.44^{+0.29}_{-0.24}$\\
270.0&50.0&$10.89^{+0.38}_{-0.29}$\\
322.5&55.0&$11.60^{+0.74}_{-0.45}$\\
\hline
\end{tabular}
\end{table}

\section{Conclusions}
\label{sec:conclusions}

We have investigated the use of read-out streaks in obtaining
photometric measurements of stars which are too bright for normal
aperture photometry in MCP-intensified CCD detectors, and in
particular the {\em Swift} UVOT. Our study is based on UVOT v-band
measurements of stars in the Tycho-2 catalogue. We find that through
the use of read-out streaks, photometric measurements can be obtained
for stars up to 2.4 magnitudes brighter than the usual coincidence
loss limit. The read-out streaks, which are formed during the frame
transfer on the CCD, are not affected by coincidence loss arising in
the CCD. Instead, we find that coincidence-loss associated with the
recharge time of the pores of the MCPs in the intensifier becomes
significant. We calibrated the MCP coincidence loss using the Tycho-2
measurements. Our analysis of the photometric scatter in the
read-out-streak measurements of Tycho-2 stars indicates that
systematics and un-accounted sources of error limit the photometric
precision to 0.1 mag over the factor 10 dynamic range for which the
read-out streaks are useful.

\section*{Acknowledgments}
\label{sec:acknowledgments}

This work was supported by the United Kingdom Space Agency (UKSA). We thank 
Rhaana Starling and Julian Osborne for useful discussions.

\bibliographystyle{mn2e}

\begin{thebibliography}{}

\bibitem[Barthelmy et~al.(2005)]{barthelmy05}
Barthelmy S. D., et~al., 2005, Space Sci. Rev., 120, 143

\bibitem[Breeveld et~al.(2010)]{breeveld10}
Breeveld A. A., et~al.,
2010, MNRAS, 406, 1687

\bibitem[Breeveld et~al.(2011)]{breeveld11}
Breeveld A. A., Landsman W., Holland S. T., Roming P., Kuin N. P. M., 
Page M. J., 
2011, AIP conf. proc. 1358, 373

\bibitem[Burrows et~al.(2005)]{burrows05}
Burrows D. N., et~al., 2005, Space Sci. Rev., 120, 165

\bibitem[Eberhardt(1981)]{eberhardt81}
Eberhardt E.H., 
1981, IEEE Transactions on Nuclear Science, NS-28, 712

\bibitem[ESA(1997)]{esa97}
ESA, 1997, The Hipparcos and Tycho Catalogues, ESA SP-1200

\bibitem[Fordham et~al.(1989)]{fordham89}
Fordham J. L. A., Bone D. A., Read P. D., Norton T. J., 
Charles P. A., Carter D., Cannon R. D., Pickles A. J., 
1989, MNRAS, 237, 513

\bibitem[Fordham et~al.(2000a)]{fordham00a}
Fordham J. L. A., Kawakami H., Michel R. M., Much R., Robinson J. R.,
2000a, MNRAS, 319, 414

\bibitem[Fordham, Moorhead \& Galbraith(2000b)]{fordham00b}
Fordham J. L. A., Moorhead C. F. \& Galbraith R. F., 
2000b, MNRAS, 312, 83

\bibitem[Gehrels et~al.(2004)]{gehrels04}
Gehrels N., 2004, ApJ, 611, 1005

\bibitem[H{\o}g et~al.(2000)]{hog00}
 H{\o}g E., et~al., 2000, A\&A, 355, L27

\bibitem[Johnson \& Morgan(1951)]{johnson51}
Johnson H. L. \& Morgan W. W., 1951, ApJ, 114, 522

\bibitem[Kawakami et~al.(1994)]{kawakami94}
Kawakami H., Bone D., Fordham J., Michel R.,
1994, Nuclear Instruments and Methods in Phys. Res. A, 348, 707

\bibitem[Kuin \& Rosen(2008)]{kuin08}
Kuin N. P. M. \& Rosen S. R., 
2008, MNRAS, 383, 383

\bibitem[Maccacaro et~al.(1988)]{maccacaro88}
Maccacaro T., Gioia I.M., Wolter A., Zamorani G., Stocke J.T., 
1988, ApJ, 326, 680

\bibitem[Mason et~al.(2001)]{mason01}
Mason K. O., et~al.,
2001, A\&A, 365, L36

\bibitem[Massa \& Fitzpatrick(2000)]{massa00}
Massa D. \& Fitzpatrick E. L., 
2000, ApJS, 126, 517

\bibitem[Pickles(1998)]{pickles98}
Pickles A. J., 1998, PASP, 110, 863

\bibitem[Poole et~al.(2008)]{poole08}
Poole T. S., et~al.,
2008, MNRAS, 383, 627

\bibitem[Racusin et~al.(2008)]{racusin08}
Racusin J. L., et~al.,
2008, Nature, 455, 183

\bibitem[Roming et~al.(2005)]{roming05}
Roming P. W. A., et~al.,
2005, Space Science Reviews, 120, 95

\bibitem[Skrutskie et~al.(2006)]{skrutskie06}
Skrutskie M. F., et~al., 2006, AJ, 131, 1163

\bibitem[Talavera(2011)]{talavera11}
Talavera A., 2011, 
Technical Report XMM-SOC-CAL-TN-0019 issue 6.0, {\it XMM-Newton} 
Optical and UV monitor (OM) Calibration Status, ESA;
http://xmm2.esac.esa.int/docs/documents/CAL-TN-0019.pdf

\end{thebibliography}

\appendix
\vspace{3mm}
\noindent
{\bf APPENDIX A: recommended procedure for obtaining photometry from 
read-out streaks}

\begin{enumerate}
\item{Mask the raw image from bright sources, and mask fainter sources on a 
column by column basis so that the read-out streaks are preserved (see 
Section~\ref{sec:measurement} for more detailed recommendations regarding 
masking).} 
\item{Measure the mean counts per non-masked pixel in each column and 
scale to a 16 row aperture.}
\item{Identify the columns containing the read-out streak of interest 
and measure its count rate in a 16 pixel aperture, subtracting a suitable 
background estimate. Estimate the uncertainty on the count-rate 
assuming Poisson statistics.}
\item{Correct the count rate for coincidence loss using 
Equation~\ref{eq:coicorr} with $t_{MCP}$=0.236~ms and the appropriate value 
of $S$ from Table~\ref{tab:windowmodes}. Scale the 
uncertainty by the same factor as the count rate.}
\item{Check that the coincidence-loss corrected count rate is below the 
maximum recommended count rate given in Table~\ref{tab:windowmodes}, and 
the star is not heavily saturated (see Figure~\ref{fig:saturated}). We only 
recommend that photometry is derived from the read-out streak 
if these conditions are satisfied.} 
\item{Apply the large-scale sensitivity correction according to the 
position of the star within the
image\footnote{In the calibration database (CALDB) 
used by the standard {\em Swift} {\sc ftools} 
(http://heasarc.gsfc.nasa.gov/docs/heasarc/caldb/data/swift/uvota)
this correction is contained 
within the file swulssens20041120v003.fits (where v003 refers to the 
version number and will be incremented as the calibration is updated).}, 
and the time-dependent 
sensitivity correction\footnote{In the calibration database (CALDB) 
used by the standard {\em Swift} {\sc ftools} this correction is contained 
within the file swusenscorr20041120v003.fits (where v003 refers to the 
version number and will be incremented as the calibration is updated).}
 appropriate to the time of the UVOT observation 
\citep{breeveld10} to the coincidence-loss corrected count rate and 
its uncertainty.}
\item{Obtain a magnitude and associated uncertainty from the 
coincidence-loss corrected count-rate and 
its uncertainty using
\[
M = ZP -2.5 \log_{10} R_{i}
\]
 where $M$ is the 
magnitude, $R_{i}$ is the coincidence-loss corrected count rate 
and ZP is the zeropoint
 given in Table~\ref{tab:zeropoints} for the relevant UVOT passband and 
window mode.}
\item{Add in quadrature a systematic uncertainty of 0.1 mag to the 
statistical uncertainty on the magnitude.}
\end{enumerate}

\vspace{3mm}
\noindent
{\bf APPENDIX B: potential extension to XMM-OM}

\vspace{3mm}
\noindent
The principles for obtaining read-out-streak photometry outlined in
this paper are expected to apply equally to XMM-OM, though we caution
that we have not verified experimentally any aspect of read-out-streak
photometry with XMM-OM. XMM-OM is operated with different frame times
to UVOT, but the frame-transfer time is the same, so the ratio of
exposure times of the static image to the frame transfer $S$ is easily
calculated for the XMM-OM frame times. Using $S$, and the calibrated
zeropoints for normal aperture photometry \citep{talavera11},
zeropoints appropriate for read-out streak photometry with XMM-OM can
thus be derived in an identical fashion to the zeropoints presented
here for UVOT. The only experimentally-determined parameter in the
transformation of UVOT read-out streak measurements to photometry is
the effective recharge time of the MCP, $t_{MCP}$. This timescale is
expected to depend on the product of the recharge resistance and 
the capacitance per channel
of the final-stage
MCP, which in turn will depend on its physical characteristics
such as material, dimensions and pore layout \citep{eberhardt81}. 
The MCPs used for the
UVOT and XMM-OM have identical specifications and were procured in the
same time period from the same manufacturer, the UVOT MCPs being
spares from the development of XMM-OM. Therefore our expectation is
that the effective recharge timescale of the XMM-OM MCP should be very
similar to the $2.36\times 10^{-4}$~s measured for the UVOT MCP.

\label{lastpage}
\end{document}